\documentclass[aps,pra,twocolumn,nopacs,longbibliography,
  numerical,floatfix]{revtex4-2}
\usepackage{epsfig}
\usepackage{graphicx}
\usepackage{dcolumn}
\usepackage{amsthm,amsmath,amsfonts}
\usepackage{mathtools}
\usepackage{comment}
\usepackage[colorlinks]{hyperref}
\usepackage{xcolor}
\usepackage{orcidlink}
\usepackage{rotating}
\usepackage{cancel}
\usepackage{orcidlink}
\usepackage[normalem]{ulem}
\usepackage{multirow}
\usepackage{epstopdf}
\AppendGraphicsExtensions{.pdf}
\usepackage[normalem]{ulem}
\usepackage{xcolor}
\newcommand{\redsout}{\bgroup\markoverwith{\textcolor{red}{\rule[0.5ex]{2pt}{0.4pt}}}\ULon}
\usepackage{epstopdf}
\AppendGraphicsExtensions{.pdf}

\begin{document}
 \title{Characterization of Thermalization Behaviour in a Generalized Aubry-Andr\'e Model}
\author{S. Mal\orcidlink{https://orcid.org/0000-0002-9576-6732}$^{{\dagger}}$}
\email{Email: subhankamal@gmail.com}
\author{D. K. Nandy\orcidlink{https://orcid.org/0000-0002-1667-4347}$^{{\ddagger}}$} 
\email{Email: nandy.pawan@gmail.com}
\author{B. K. Sahoo\orcidlink{https://orcid.org/0000-0003-4397-7965}$^{{\dagger}}$}

\affiliation{$^{\dagger}$Atomic, Molecular and Optical Physics Division, Physical Research Laboratory, Navrangpura, Ahmedabad 380009, India \\
$^{\ddagger}$P. G. Department of Physics, S. K. C. G. (Auto.) College, Paralakhemundi, Odisha 761200, India}

\begin{abstract}
Although random matrix theory provides a fundamental framework for characterizing quantum chaos, encompassing both ergodic and localized phases, a comprehensive understanding of the universal features governing the critical transition remains elusive in many disordered and quasi-random systems. In this study, we explore the ergodic-to-many-body localization transition in the generalized Aubry-André model with interacting spinless fermions. Using the concept of Frobenius norm of an adiabatic gauge potential, we construct a phase diagram that captures the sensitivity of the eigenspectrum to infinitesimal adiabatic gauge deformations. To examine the stability of the critical disordered strength with respect to system size, we perform an unbiased finite-size scaling analysis via cost-function minimization techniques. Additionally, by analyzing the adjacent gap ratio and spectral form factor, we determine the scaling behavior of the Thouless time as a function of the disorder strength.
\end{abstract}

\maketitle

\section{Introduction}

Following the pioneering work of Anderson \cite{Anderson_PR:1958}, the investigation of localization in quantum systems has remained a central focus in condensed matter physics for many decades. It is now widely recognized that localization phenomena extend beyond systems with quenched random disorder, encompassing a broader range of physical contexts and mechanisms. For example, quasi-periodic systems constitute a unique class of deterministic, non-random models that display localization phenomena along with a range of related effects. These include the presence of mobility edges, stable multifractal eigenstates, coexistence of extended and localized states, anomalous transport behaviors, and, in presence of interaction, the emergence of many-body localized (MBL) phases \cite{Kohmoto_PRL:1983, Prange_PRB:1983, Das_Sarma_PRL:1988, Boers_PRA:2007, Biddle_PRA:2009, Lahini_PRL:2009, Biddle_PRL:2010, Ganeshan_PRL:2015, Morales_PRA:2014, Li_PRL:2015, Modak_NJP:2016, Purkayastha_PRB:2017, Purkayastha_PRB:2018, Sthitadhi_SciPost:2018, Luschen_PRL:2018}. 
From a theoretical standpoint, the MBL in interacting many-body systems is significantly more intriguing in quasi-periodic systems than in disordered ones, largely because of the successful experimental realization of these models in numerous cold-atom experiments \cite{Schrieber_Sci:2015, Bordia_NatPhys:2017, Luschen_PRL:2017, Bordia_PRX:2017}. 

Over the past few decades, disordered systems have been extensively studied theoretically across many one-dimensional and quasi-one-dimensional models. Due to their inherently random nature, averaging over multiple disorder realizations is essential to accurately characterize their properties. In contrast, for quasi-periodic models, although averaging remains important, analysis typically focuses on the eigenvalues and eigenstates within a single realization to precisely understand system behavior. The Aubry-André (AA) model, which is the simplest quasi-periodic model, exhibits a localization transition in one dimension, and various extensions of this model predict the presence of mobility edges in the eigenvalue spectrum \cite{Ganeshan_PRL:2015, Li_PRL:2015, Yao_PRL:2019, Liu_SciPost:2022}. Generally, quasi-periodic systems are fundamentally different from disordered systems \cite{Khemani_PRL:2017}, as the potential in the former is deterministic with correlated structure, whereas the latter is characterized by uncorrelated random disorder. 

Throughout the development of the field, random matrix theory \cite{Brody_RMP:1981, Guhr_PhysRep:1998} has proven remarkably effective in understanding quantum chaos. It has been theoretically established that chaotic quantum systems exhibit level repulsion, with their spectral statistics belonging to one of three universality classes depending on the symmetry of the Hamiltonian: the Gaussian orthogonal ensemble (GOE), the Gaussian unitary ensemble, or the Gaussian symplectic ensemble  \cite{Wigner_AM:1957, Dyson_JMP:1962, Mehta_Book}. In contrast, localized systems do not display level repulsion and typically follow a Poissonian distribution of level spacings \cite{Berry_PRSA:1977}. These correlations among energy levels can be effectively characterized through the distribution of adjacent level spacings. In this context, delocalization and localization are often investigated either via spectral diagnostics \cite{Pal_PRB:2010, Kjall_PRL:2014} or through matrix elements of some local or global observables evaluated using the eigenstates \cite{Pal_PRB:2010, Khatami_PRE:2012, Serbyn_PRX:2015, Serbyn_PRB:2017, Panda_EPL:2019, Oganesyan_PRB:2007, Suntajs_PRB:2020, Serbyn_PRB:2016, Buijsman_PRL:2019, Sierant_PRB:2019}. For example, the thermalization timescale, known as the Thouless time ($t_{\rm Th}$), can be extracted across parameter space using the spectral form factor (SFF), which is the Fourier transform of the two-point correlation function in the eigenvalue spectra \cite{Suntajs_PRE:2020, Prakash_PRR:2021, Das_PRR:2025}. While the analysis of the SFF offers a nuanced understanding of the thermalization process, precise characterization of the transition point from the ergodic to the MBL phase remains an open challenge. In particular, the stability of the phase boundary with respect to system size is not yet well understood through these diagnostics \cite{Suntajs_PRE:2020}.

\begin{figure}[t]
\includegraphics[width=8cm,height=3cm]{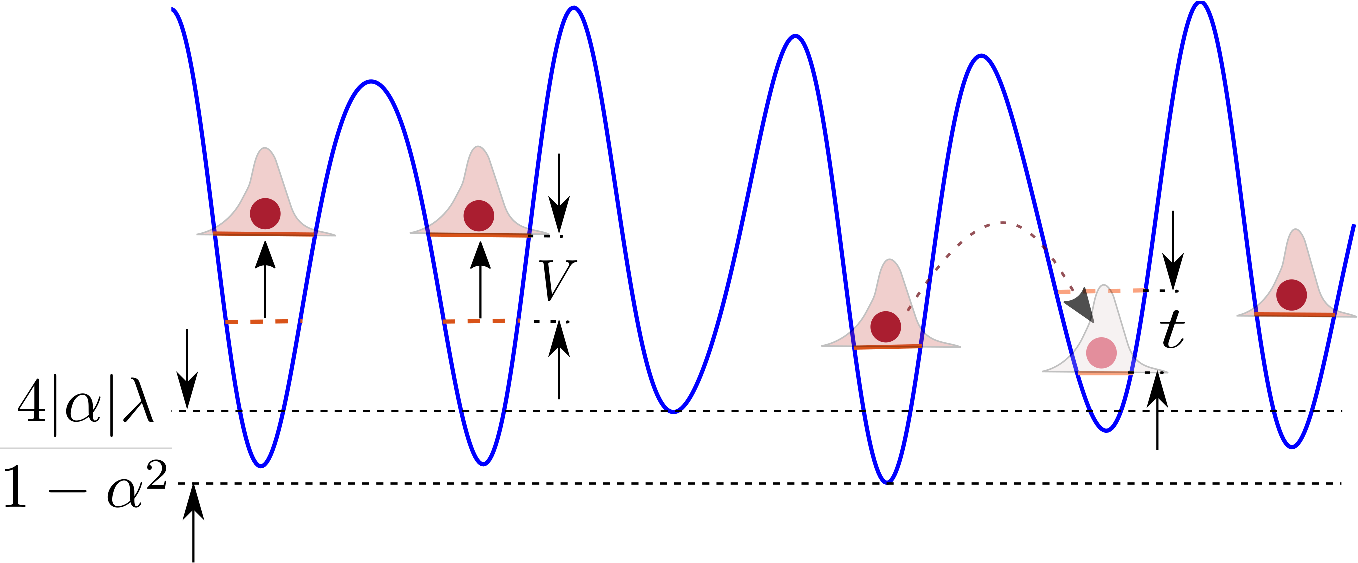}
\caption{A schematic representation of the generalised Aubry- Andr\'{e} model. The maxima and minima of the quasi-periodic potential of the considered system are governed by two parameters $\alpha$ and $\lambda$, with the quasi-periodic wavenumber given by the irrational number $q = (\sqrt{5} -1)/2$. Here, interaction between the spinless fermions is considered only when there are two successive sites occupied.}
\label{GAApot}
\end{figure}

An alternative approach to probe the intriguing nature of quantum chaos involves examining the sensitivity of eigenstates under adiabatic deformations of the system Hamiltonian \cite{Pandey_PRX:2020, Sels_PRE:2021}. It has been demonstrated that the Frobenius norm of the adiabatic gauge potential (AGP), or equivalently the fidelity susceptibility $(\chi_n )$, $n$ denotes a general eigenstate index \cite{Sels_PRE:2021}, is highly responsive to the change in the disorder parameter value. Indeed, $\chi_n$ exhibits exponential scaling in the chaotic regime and polynomial scaling in the MBL regime as a function of system size. Importantly, this quantifier involves both eigenvalues and eigenstates of the unperturbed Hamiltonian, distinguishing it from conventional diagnostics. Recently, the AGP-based analysis has been applied to a mass-deformed Sachdev–Ye–Kitaev model and a class of random matrix models to investigate their localization properties through numerical diagonalization \cite{nandyPRB2022, CadezNJP2024}. While this method has been explored extensively for disordered spin systems, all-to-all models, and random matrix models, a systematic analysis to characterize the MBL transition for the finite-sized systems and its behavior in the thermodynamic limit in a quasi-periodic model—such as the generalized AA (GAA) model \cite{Ganeshan_PRL:2015}, which features a single-particle mobility edge—remains unexplored using the AGP framework. 

Motivated by this AGP (or equivalently, fidelity susceptibility) diagnostics introduced for probing MBL, we investigate the GAA model to understand its thermalization behavior and the nature of its phase transition using such novel techniques. While the AA model and its generalized variants have been studied through conventional MBL quantifiers \cite{Li_PRL:2015}, key questions regarding the stability of the phase boundary in the thermodynamic limit and the behavior of the correlation length at the transition point remain unresolved. Specifically, we investigate the following issues: (i) the impact of finite-size effects on the MBL transition in the GAA model; (ii) the sensitivity of the AGP-based fidelity susceptibility relative to conventional indicators of the transition; and (iii) the scaling behavior of the finite-size critical disorder.  To address these issues, we carried out a systematic numerical analysis of the scaled fidelity susceptibility for two types of gauge-deforming potentials: local and extensive. Our results reveal that the scaled $\chi_n$, as a function of disorder strength, exhibits three distinct regimes of sensitivity to these gauge potentials. Qualitatively, the resulting phase diagram resembles that of the random $XXZ$ model \cite{Suntajs_PRE:2020}, the sensitivity characteristics vary significantly across different regions of the phase diagram. To assess the stability of the critical disorder, we perform a comprehensive data collapse analysis of the scaled $\chi_n$ and the adjacent gap ratio, employing cost-function minimization techniques. In addition, we estimate thermalization time scales in the presence of a mobility edge by examining the SFF, following the approach outlined in Refs. \cite{Edward_JPC:1972, Schiulaz_PRB:2019, Suntajs_PRE:2020}. Our findings indicate that the dependence of $t_{\rm Th}$ on the GAA parameter is substantial and can even exceed the Heisenberg time ($t_H$), exhibiting behavior analogous to that observed in the strong disorder regime.

\section{The model}\label{sec.2}

The Hamiltonian for the spinless fermionic particles with nearest-neighbor interactions in the GAA model considered in our analysis is given by 
\begin{eqnarray}
H = -t\sum_{\langle i j\rangle} a_i^\dagger a_j +  \lambda \sum_i C_i n_i  + V\sum_{\langle i j\rangle} n_in_j ,
    \label{eq1}
\end{eqnarray}
where $a_i$ ($a_i^{\dagger}$) are fermionic annihilation (creation) operators satisfying the anti-commutation relation  $\{a_i, a_j^{\dagger}\} = \delta_{ij}$, and $n_i = a_i^{\dagger}a_i$ is the number operator at the $i^{th}$ site. In the above Hamiltonian, the first term represents the kinetic energy with a uniform tunneling amplitude $t$ between nearest-neighbor sites, denoted by $\langle i j \rangle$. The second term accounts for the on-site potential energy characterized by a site-dependent quasi-periodic modulation coefficient $C_i$ with strength $\lambda$. The third term describes nearest-neighbor interactions between fermions with a constant interaction strength $V$. Unless stated otherwise, we set $t=1$ as our energy unit.

In the standard AA model, $C_i$ is defined as $C_i = \cos(2\pi q i + \phi)$, where the irrational wavenumber $q$ introduces the quasi-periodic behavior in the potential, and $\phi$ is a random offset parameter in the phase factor. In our calculations, $\phi$ is treated as a random variable to obtain the disorder-averaged physical quantities, although it can be tuned experimentally. However, in the present GAA model, we consider $C_i = \frac{2\cos(2\pi q i + \phi)}{1 - \alpha \cos(2\pi q i + \phi)}$, where $\alpha$ is a generalized AA parameter. Notably, when $\alpha = 0$, the GAA model reduces to the standard AA model. For convenience, we restrict $\alpha$ to the range $[0, 1)$. Figure \ref{GAApot} provides a schematic illustration of the GAA potential described by Eq. (\ref{eq1}) for clarity. As shown, the maxima and minima of the quasi-periodic potential depend on the parameters $\alpha$ and $\lambda$, with the quasi-periodicity determined by an irrational number $q=(\sqrt{5} - 1)/2$. 

Additionally, we impose U(1) symmetry by conserving the total particle number throughout the analysis; i.e. $\sum_i n_i = N$. In the non-interacting limit, the AA model ($\alpha = 0$) exhibits a self-duality condition at the critical disorder strength $\lambda^* = t $, marking the transition between extended and localized phases for $\lambda < \lambda^*$ and $\lambda^* > \lambda$, respectively. The non-interacting GAA model ($\alpha \neq 0$) show a mobility edge described by the relation $\alpha E = 2 \textrm{sgn}(\lambda)(|t| - |\lambda|)$ \cite{Ganeshan_PRL:2015}, where $E$ denotes the ordered energy spectrum.

\begin{figure}[t]
\includegraphics[width=8.6cm,height=5cm]{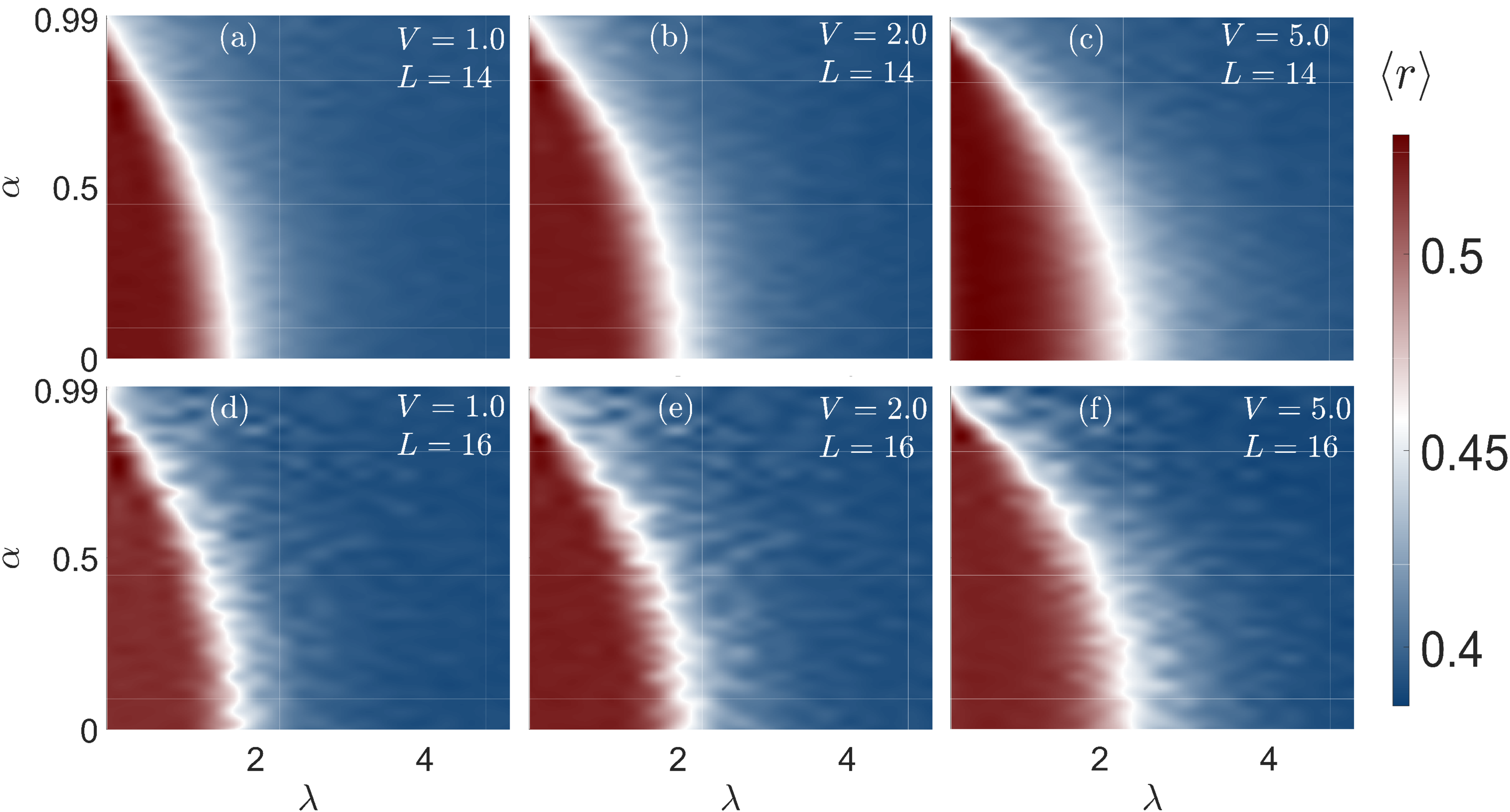}
\caption{The average of adjacent gap ratio $\langle r \rangle$ defined in Eq. (\ref{eq2}) is presented in the plane of $\alpha$ and $\lambda$ for different values of interaction with $L = 14$ (upper row) and $L=16$ (lower row). For a fixed value of $\alpha$, we observe $\langle r \rangle \approx 0.53$ for smaller values of $\lambda$, signifying that the level statistics follow the GOE distribution.}
\label{fig 2}
\end{figure}
\begin{figure*}[t]
\includegraphics[width=15cm,height=12cm]{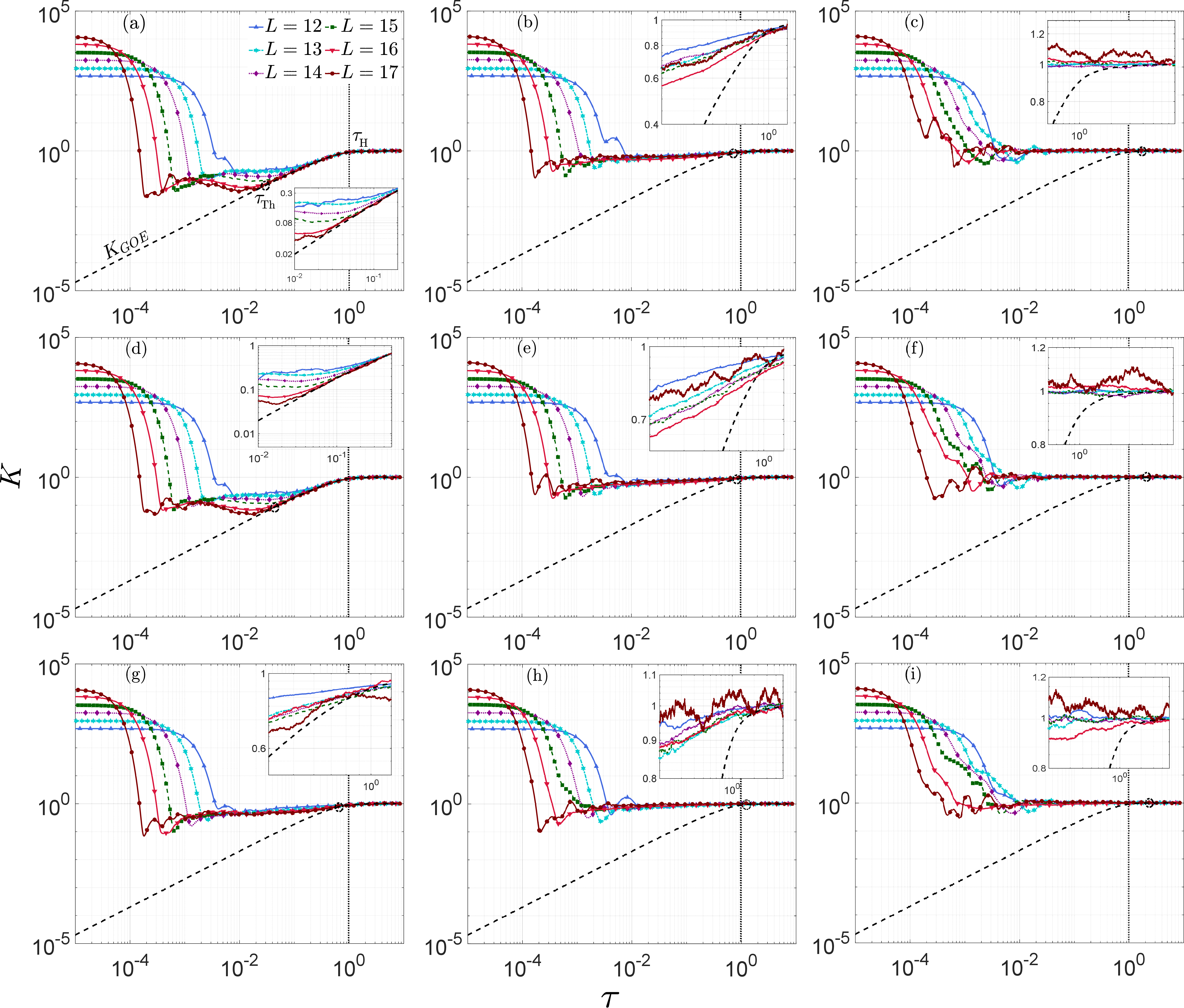}
\caption{Spectral form factor $K(\tau)$ is shown as a function of scaled time $\tau$ for different values of $\alpha = 0$ (a, b, c), $0.3$ (d, e, f), $0.7$ (g, h, i) and $\lambda = 0.65$ (a, d, g), $1.35$ (b, e, h), $10.15$ (c, f, i). The black dashed line corresponds to $K_{GOE}(\tau)$ and the dotted vertical line represents the scaled Heisenberg time $\tau_H$. The Thouless time $\tau_{Th}$ is identified as $K\rightarrow K_{GOE}$ and indicated by the dashed circles. For a lower value of $\lambda$, the ratio of Thouless time and Heisenberg time $\tau_{Th}/\tau_H << 1$ for all three values of $\alpha$. As the value of $\lambda$ increases, the ratio approaches unity and for large $\lambda$ goes past unity. The insets represent the magnified images near Thouless time.}
\label{fig 3}
\end{figure*}

\section{Numerical Approaches}\label{sec.NuAp}

To investigate the correlation among the energy values and their spectral properties for different system sizes, we employ the exact diagonalization (ED) method for system sizes $L \leq 18$. Specifically, we perform numerical calculations for lattice sizes $L = 12$, $13$, $14$, $16$, $17$ and $18$. For these systems, we consider the filling fraction to be exactly half-filling ($N = L/2$) when $L$ is even, and nearly half-filling ($N = (L + 1)/2$) when $L$ is odd. The corresponding Hilbert space dimensions $\mathcal{D}$ are: $924, 1716, 3432, 6435, 12870, 24310$, and $48620$ for $L = 12$, 13, 14, 15, 16, 17 and 18, respectively. This demands carrying out ED using a high-performance computing facility. 

We analyze the entire energy spectrum to compute the adjacent level spacing ratio ($r$) across the full range of $\alpha$ and $\lambda$ parameters discussed in the previous section. To obtain statistically reliable results, we average over multiple realizations: for $L=14$ and $16$, we use $N_{\rm sample} = 200$ and $50$, respectively, by sampling different random offsets $\phi$. For the SFF calculations, we adopt a similar approach up to $L=17$, increasing the number of samples for a set of smooth data. Specifically, for $L=12, 13, 14, 15, 16$ and $17$, we use $N_{\rm sample} = 2000$, $1500$, $1000$, $200$, $100$, and $50$, respectively. Finally, we evaluate $\chi_n$ for various system sizes considering a microcanonical energy shell as has been done in \cite{Pandey_PRX:2020, Suntajs_PRE:2020}. For $L \leq 16$, we focus on a subset of interior eigenstates with eigenvalues between $-0.1$ and $0.1$ of a scaled Hamiltonian $\tilde{H}$, which has eigenvalues in the range $\tilde{E} \in [-1,1]$. For $L=18$, we employ the Chebyshev filtering scheme \cite{Pieper_JCP:2016} to extract approximately $1500$ eigenstates centered around the middle of the spectrum for calculating an average value of $r$ and $\chi_n$.

\section{Results and Discussion}

Since our main objective in this study is to characterize the onset of the MBL transition in the presence of a mobility edge within the GAA model, we first analyze the $r$ values to investigate the level statistics associated with the transition. Subsequently, we examine the long-range spectral correlation to evaluate the thermalization behavior with respect to the various system parameters. To end, we discuss the fate of the transition point in the thermodynamic limit by calculating $\chi_n$ and analyzing its finite-size scaling behavior. A comprehensive discussion of these properties is provided in the following sections.

\subsection{Adjacent gap ratio}

To identify different many-body phases of the GAA model, one needs to examine the many-body eigenspectrum. For this analysis, we calculate the adjacent gap ratio for the $i^{th}$ eigenstate defined as
\begin{eqnarray}
r_i = \frac{{\rm min}(\delta_{i+1},\delta_i)}{{\rm max}(\delta_{i+1},\delta_i)} ,
\label{eq2}
\end{eqnarray}
where $\delta_i = E_{i+1} - E_i$ stands for successive gaps with the ordered energy spectrum $\{E_i\}$. For chaotic systems, the average value of the adjacent gap ratio  ($\langle r\rangle$) over samples and the number of eigenvalues can be computed by considering an appropriate random matrix ensemble. For example, GOE, which is appropriate for systems with time-reversal symmetry, can give $\langle r\rangle \approx 0.53$, while a Poissonian level spacing distribution that is applicable for the MBL systems can give $\langle r\rangle = 2 {\rm ln}(2) -1 \approx 0.39$ \cite{Mehta_Book}. 

In Fig. \ref{fig 2}, we present phase diagrams of the average gap ratio ($\langle r \rangle$) in the $\alpha$ and $\lambda$ parameter plane for three different interaction strengths $V = 1$, $2$ and $5$, considering system sizes $L = 14$ and $16$. From the figure, it is evident that for $\alpha = 0$ and $V = 1$, the system undergoes a transition from an ergodic to a non-ergodic phase within the disorder range $1.4 \leq \lambda \leq 1.9$, indicating the critical disorder $\lambda^*$ within this range. It is also observed that the $\lambda^*$ value increases with increasing interaction strength \cite{Lev_PRL:2015}, as shown for $V = 2$ and $V = 5$ in the same figure. Additionally, we find that as the value of $\alpha$ increases, the finite-size critical disorder starts to decrease and eventually vanishes as $\alpha$ approaches unity. This trend is particularly notable as it indicates that increasing $\alpha$ suppresses ergodicity, thereby favoring MBL and making the system more susceptible to localized phases. 

\begin{figure}[h]
\includegraphics[width=8.5
cm,height=2.8cm]{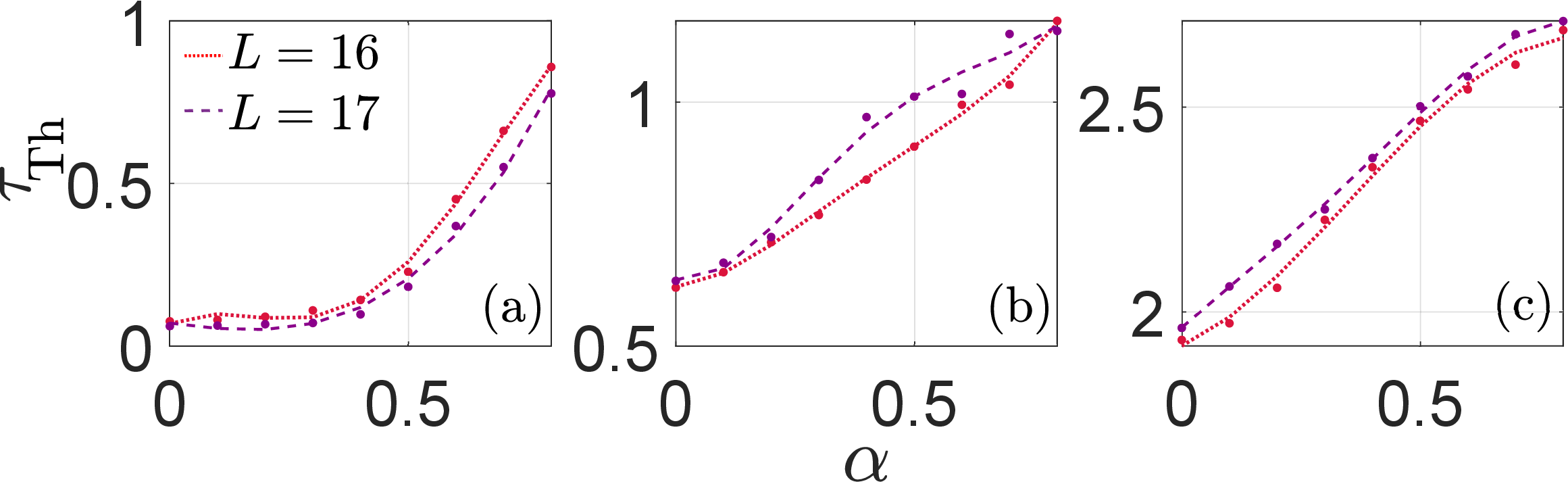}
\caption{The scaled Thouless time $\tau_\textrm{Th}$ is shown as a function of $\alpha$ for three different values of $\lambda = 0.65$ (a), $1.35$ (b) and $10.15$ (c) with $L = 16$ and $17$. In the ergodic regime, with $\lambda < 1$, $\tau_\textrm{TH}$ increases rapidly as $\alpha$ approaches unity.}
\label{fig ThAlpha}
\end{figure}

\begin{figure*}[t]
\includegraphics[width=16cm,height=5cm]{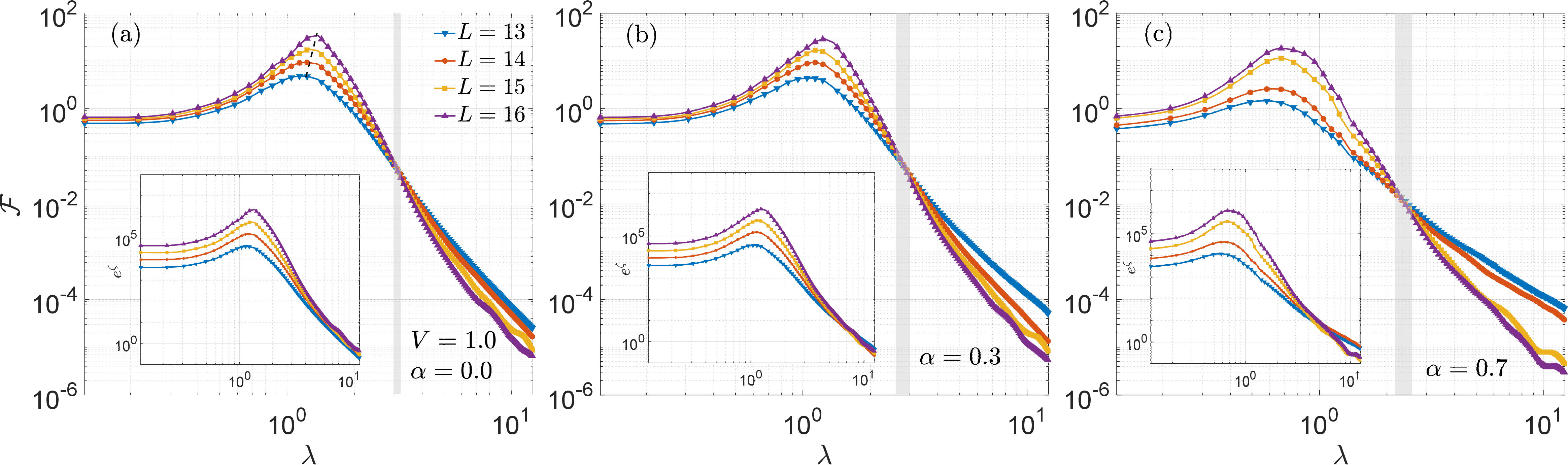}
\caption{Scaled fidelity susceptibility $\mathcal{F}$ for the local operator $n_{L/2}$ ($n_{(L+1)/2}$) for even (odd) lattice sizes as a function of $\lambda$ is plotted for different values of $\alpha$. We consider $L = 13, 14, 15$ and $16$ for $\alpha = 0.0, 0.3$ and $0.7$. In the insets, we show the behavior of unscaled fidelity susceptibility $e^{\zeta}$. The crossing of the three curves, indicated with the shaded region, specifies the transition $\lambda^*$ between ergodic and localized phases. With increasing $\alpha$, $\lambda^*$ decreases as predicted by the earlier analysis. The dashed line indicates the peak-drift behavior towards higher values of $\lambda$ as the system size is increased.}
\label{AGPlocal}
\end{figure*}

\subsection{Spectral Form Factor}\label{sec.3}

The analysis of $\langle r \rangle$ reflects the level repulsion between different adjacent energy levels and thereby demonstrates only short-range spectral correlations. However, it is insufficient to reveal important physical insights, as it lacks information on long-range correlations in the spectrum. Furthermore, the phase diagram obtained using the $\langle r \rangle$ values does not give a concrete idea about the characteristics of the correlation length in detail and the value of critical disorder strength for the transition from ergodic to the MBL phase.

The average level spacing $\delta$ ($\bar{\delta}$) typically decays exponentially with the system size $L$ and its inverse, which is related to the Heisenberg time, $t_H$ as $t_H = \hbar/\bar{\delta}$, increases exponentially with $L$. The knowledge of $t_H$ is useful as it indicates the time after which the system can not self-thermalize. On the other hand, $t_{\rm Th}$ indicates the time required for the system to thermalize \cite{Sierant_PRL:2020}. To quantify the thermalization properties with respect to the aforementioned system parameters, a more comprehensive diagnosis of SFF \cite{Suntajs_PRE:2020} is needed. The SFF, which is nothing but the Fourier transform of the two-point spectral correlation, is defined as
\begin{eqnarray}
K(\tau) = \frac{1}{Z}\left\langle\left|\sum_{\alpha = 1}^{\mathcal{D}}\rho(\epsilon_\alpha)e^{-i2\pi \epsilon_\alpha \tau}\right|^2\right\rangle ,
\end{eqnarray}
where $Z$ is the normalization factor, $\{\epsilon_1 \le \epsilon_2 \le . . . \le \epsilon_{\mathcal{D}}\}$ are the complete set of unfolded eigenvalues with the Hilbert space dimension $\mathcal{D}$ and $\tau$ is the scaled time in the unit of $t_H$, making the scaled Heisenberg time $\tau_H = 1$, and $\rho(\epsilon)$ is the Gaussian filter function. Here, the normalization factor $Z$ is chosen such that $K(\tau) \rightarrow 1$ when $\tau \rightarrow \infty $. The Gaussian filter function is given by $\rho(\epsilon) = {\rm exp}\left[- \frac{(\epsilon - \bar{\epsilon})^2}{2(\eta\Gamma)^2}\right]$, which helps to smooth out the unfolded spectrum. In this function, $\bar{\epsilon}$ is the average unfolded energy $\bar{\epsilon}$ and $\Gamma^2$ is the variance. Furthermore, the dimensionless parameter $\eta$ is used to control the fraction of the eigenvalues that can be included to calculate $K(\tau)$. For our calculations, we consider $\eta = 0.5$, however, the value $K(\tau)$ is less dependent on the value of $\eta$ as has been shown for the random model \cite{Suntajs_PRE:2020}. The detailed procedure to obtain the unfolded energy is given in Ref. \cite{Suntajs_PRE:2020}. The unfolding procedure essentially removes the non-universal average density of states to isolate the universal fluctuations present in the raw energy levels. The dimensionless unfolded energy level obtained in this way sets the mean level spacing as unity. A typical approach to analyze quantum chaotic dynamics is to extract the ratio $\tau_{\rm Th}/\tau_H \equiv t_{\rm Th}/t_{H}$ from the $K(\tau)$ curve that may serve as an indicator of ergodicity breaking in finite systems.

\begin{figure*}[t]
\includegraphics[width=16cm,height=5cm]{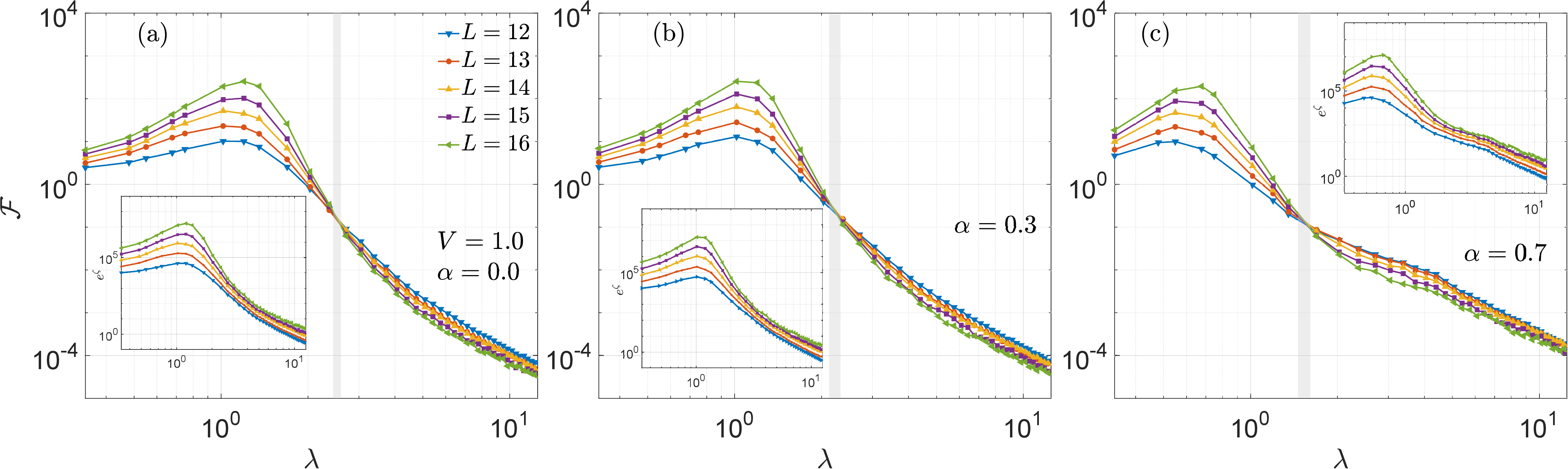}
\caption{Scaled fidelity susceptibility $\mathcal{F}$ for the extended AGP operator ($\sum_{i}n_i n_{i+1}$) is plotted as a function of $\lambda$ for $\alpha = 0$ (a), $0.3$ (b) and $0.7$ (c) with $L = 12, 13, 14, 15$ and $16$. Insets present the unscaled fidelity $e^\zeta$. The gray shaded region in the main plots defines the crossing of $\mathcal{F}$ for different system sizes.}
\label{AGPglobal}
\end{figure*} 

In Fig. \ref{fig 3}, we plot $K(\tau)$ for three different values of $\lambda$ such as $\lambda = 0.65, 1.35$ and $10.15$ and for the GAA parameter $\alpha = 0.0$, 0.3, and 0.7. This set of parameters is chosen in such a way that the three values of $\lambda$ reside in three different regimes (ergodic, intermediate and localized) of the phase diagram shown in Fig. \ref{fig 2}. Unless stated otherwise, we have fixed the interaction strength $V = 1$ henceforth in our calculations, as the focus is to study the thermalization properties and investigate the MBL transition with respect to the other system parameters in the presence of a fixed nearest-neighbor interaction. The black dashed line represents the SFF for GOE, defined for a finite system with time-reversal symmetry, as $K_{GOE} = 2\tau - \tau {\rm ln}(1 + 2\tau)$ for $\tau < 1$ \cite{Suntajs_PRE:2020}. The scaled Thouless time is extracted by the following criterion
\begin{eqnarray}
\log_{10}\left[\frac{K(\tau_{\rm Th})}{K_{GOE}(\tau_{\rm Th})}\right] = \varepsilon
\end{eqnarray}
by making $\varepsilon$ reasonably small to avoid noisy data in $K(\tau)$. We use $\varepsilon=0.02$ for all the parameters of our model. The choice of $\varepsilon$ is made to ensure the robustness of the calculation, since the resulting data for $\log_{10}[K(\tau)/K_{{GOE}}(\tau)]$ remain smooth for values of $\varepsilon$ above this threshold.
Figs. \ref{fig 3}(a), (d) and (g), show the behavior of $K(\tau)$ for $\lambda = 0.65$ with $\alpha = 0.0, 0.3$, and $0.7$, respectively. As can be observed, for $\alpha = 0.0$, the ratio $\tau_{\rm Th}/\tau_H$ decreases with respect to system size, implying that the ergodic nature of the system is kept intact in the thermodynamic limit. Then, with the increase in the $\alpha$ value $\tau_{\rm Th}$ increases significantly and approaches the $\tau_H$ limit (see Fig. \ref{fig 3}(g)). This implies that the GAA model undergoes a transition from the ergodic to MBL phase for the GAA parameter $\alpha: 0 \rightarrow 1$. The trend is also quite visible from the phase diagram given in Fig. \ref{fig 2} for $\langle r \rangle$ in the $\lambda$ and $\alpha$ plane.  

Next, we calculate $K(\tau)$ for the disorder strength with $\lambda = 1.35$, which corresponds to the intermediate regime of the phase diagram. Its behavior is shown in the Figs. \ref{fig 3} (b), (e) and (h) for the same three representative values of $\alpha$ as discussed above. As can be seen from these plots, the behavior of  $K(\tau)$ is quite similar (shows only a slope-plateau regime) for all three $\alpha$ values. The Thouless time obtained using the criterion as stated above shows a value $\tau_{\rm Th} \sim \tau_{H}$. Since the value of $\lambda$ in this intermediate regime corresponds to the mobility edge, especially towards the larger value of $\alpha$, it would be premature to say the onset of the MBL phase. Now, for the value $\lambda = 10.15$ (non-ergodic regime), the system is expected to stay in the MBL phase, as evident from Fig. \ref{fig 2}. The $K(\tau)$ value for this $\lambda$ is plotted in Figs. \ref{fig 3}(c), (f) and (i) for the three $\alpha$ values. It is important to note that $\tau_{\rm Th}$ is always greater than $\tau_H$. In fact, the ratio increases reasonably as we increase the $\alpha$ value. This higher value of $\tau_{\rm Th}$ is also observed for the random $XXZ$ and $J_1-J_2$ models in the MBL regime \cite{Suntajs_PRE:2020} using a ergodicity breaking indicator $g = \log_{10}[\tau_H/\tau_{\rm Th}]$.

In Figs. \ref{fig ThAlpha} (a), (b) and (c), we further show the behavior of $\tau_{\rm Th}$ as a function of the GAA parameter $\alpha$ for $\lambda = 0.65$, $1.35$ and $10.15$, respectively, with system sizes $L=16$ and $17$. As can be observed in the ergodic regime, Fig. \ref{fig ThAlpha} (a) ($\lambda = 0.65$), the $\tau_{\rm Th}$ value increases significantly starting from $\alpha \sim 0.4$ and attains the $\tau_H$ limit for larger $\alpha$ value. Further, as we increase the system size, the $\tau_{\rm Th}$ value becomes smaller. With such a trend, we therefore expect in the thermodynamic limit the system to exhibit thermalization quite early. Then, for $\lambda = 1.35$ and $\lambda = 10.15$ (Figs. \ref{fig ThAlpha} (b) and (c)), the system shows a larger $\tau_{\rm Th}$ value and become higher than the $\tau_H$. This scenario merely reflects the fact that the system undergoes a phase transition from ergodic to MBL, for which the higher-order correlations (or level repulsion) in energy eigenvalues start diminishing.    

\subsection{Fidelity susceptibility}\label{sec.4}

Fairly recently, it has been observed that AGP (or its equivalent) and the fidelity susceptibility $\chi_n$ averaged over all the eigenstates can serve as another very sensitive probe to quantum chaos. To calculate the fidelity susceptibility, one needs to probe the Hamiltonian with some arbitrary parameter $\beta$ so that the effective Hamiltonian becomes $H \rightarrow H + \beta H_{1}$, where $H_{1}$ is a generic operator, perturbed part of the Hamiltonian, which does not necessarily commute with $H$. The fidelity susceptibility, for a given eigenstate $|n\rangle$ of the original Hamiltonian, is defined as \cite{Sels_PRE:2021}
\begin{eqnarray}
\chi_n = \sum_{m \neq n} \frac{|\langle n | \partial_\beta H | m \rangle|^2}{(E_n - E_m)^2} .
\label{eq LFS}
\end{eqnarray}
Indeed, the calculation of fidelity susceptibility is performed by changing the Hamiltonian adiabatically through the gauge potential $H_1$. Here, we have calculated $\chi_n$ by letting $\beta = 0$. This means we analyzed the sensitivity of the eigenstates under infinitesimal perturbation to the system. The presence of the denominator in Eq. (\ref{eq LFS}) suggests that the average $\chi_n$ can exponentially diverge in the absence of level repulsion due to rare resonance in the strong disorder regime \cite{Sels_PRE:2021}. Therefore, it is convenient to analyze the scaling behavior of $\chi_n$ by averaging its logarithm
\begin{eqnarray}
\zeta = \langle\langle {\rm ln}(\chi_n) \rangle\rangle.
\end{eqnarray}
The symbol $\langle\langle .. \rangle\rangle$ represents the average over both the number of realizations and the total number of eigenstates present over the energy shell. In Ref.  \cite{Sels_PRE:2021}, the scaling properties of $\zeta$ have been extensively studied in a random $XXZ$ model as a function of disordered strength for different system sizes. The typical characteristics of $\zeta$ for such a random model show three different scales of sensitivity with respect to the disordered parameter, namely ergodic, glassy and localized regimes \cite{Sels_PRE:2021, CadezNJP2024}.  

We now extend the above diagnostic through AGP to our present GAA model for characterizing the spectral sensitivity to adiabatic gauge deformation. In fact, we analyse the logarithmic fidelity susceptibility $\zeta$ for two types of gauge deformation potentials, such as local and global ones. To see the sensitivity of the eigenstates in the presence of a local deformation, we consider $H_{1} = n_{L/2}$ for even $L$ and $H_{1} = n_{(L+1)/2}$ for odd system sizes. In Fig. \ref{AGPlocal}, we present the fidelity susceptibility scaled with the maximum dimension of the Hilbert space, $\mathcal{F} = \exp[\zeta]/2^L$, as a function of $\lambda$ for different system sizes $L = 13, 14, 15$ and $16$. The main plots of Fig. \ref{AGPlocal}(a), (b), (c) represent scaled susceptibility $\mathcal{F}$, while the insets show unscaled fidelity susceptibility $\exp[\zeta]$ for three different values of $\alpha = 0.0, 0.3$ and $0.7$, respectively. Both the scaled and unscaled fidelity susceptibility are calculated by considering the energy window approach. As observed in the disordered spin case \cite{Sels_PRE:2021}, $\mathcal{F}$ shows rather similar asymptotic as well as intermediate behavior for the quasi-periodic GAA model. For small $\lambda$, the log-fidelity susceptibility $\zeta$ follows scaling behavior as
\begin{eqnarray}
\zeta = L \ {\rm ln}(2) + A(L) ,
\end{eqnarray}
where $A(L)$ is a suitable fitting parameter that depends on $L$. Here, unlike the disordered spin system, where $A$ is independent of system size, the quantity $A$ has slow $L$ dependence. In the intermediate values of $\lambda$, $\mathcal{F}$ has a peaked structure, signifying maximal chaotic behavior, which is drifting towards higher values of $\lambda$ with increasing system size. Such scaling signifies an exponentially long system size dependence of the relaxation time. This particular scaling behavior across different regions of the phase diagram is not readily captured from the diagnosis of the conventional quantifiers such as $\langle r \rangle$ and SFF. On the other hand, the calculation of fidelity susceptibility is computationally more rigorous, as it involves eigenstate structures in addition to the spectral statistics.
It is also important to mention that the peak values shift towards the lower values of $\lambda$ as we increase $\alpha$. Moreover, for a particular value of $\alpha$, the scaled fidelity susceptibility for different system sizes has a crossing, suggesting a signature of a phase transition \cite{Pandey_PRX:2020, Sels_PRE:2021}. Further, one can also observe that the crossing point region (shaded region in Fig. \ref{AGPlocal}) moves towards the lower value of $\lambda$ as we increase the $\alpha$ value, which is consistent with the typical behavior of the $\langle r \rangle$ phase diagram and the trend in the SFF values. This is further extensively analyzed by identifying the scaling properties of $\mathcal{F}$ with the system size near the crossing in the next subsection.

To make a comparative study, we also analyze the spectral sensitivity of the present GAA system for an extensive deformation. For simplicity, we chose the deformation operator to be $H_{1} = \sum_j n_jn_{j+1}$. In Fig. \ref{AGPglobal}, we present $\mathcal{F}$ as a function of $\lambda$ for system size $L = 12, 13, 14, 15$ and $16$. While we observe a similar behavior as in the local operator, the peak values in this case are found to be larger compared to the local case. This behavior arises because the perturbation is extensive in nature. Similar to the local operator, we see an intermediate peaked behavior followed by a crossing region (marked in the gray shaded region of the above figure) and a non-universal asymptotic tail in the non-ergodic region. For $\alpha =0$ and $0.3$, a smooth tail is observed as shown in Figs. \ref{AGPglobal}(a) and (b), respectively, in contrast to the case of $\alpha = 0.7$, shown in Fig. \ref{AGPglobal}(c). Further, it is important to note that both the local and global adiabatic gauge perturbations act as similar integrability-breaking factors to the system's extensive number of local integrals of motion \cite{Pandey_PRX:2020}, which normally protect the localized phase from thermalization. Destabilization of integrability causes level repulsion and drives the system towards quantum chaos and delocalization, independent of the spatial extent of the perturbation. As the AGP scales exponentially with system size near chaotic thresholds, it effectively washes out the spatial differences between local and global driving fields in finite-size scaling analyses.

Having discussed the typical behavior of various physical quantities such as adjacent gap ratio, SFF and AGP, we move to explore the nature of the transition at the crossing point. In particular, we analyze the type of correlation the GAA undergoes around a finite-size phase transition point. Two widely investigated hypotheses for calculating the correlation length in one-dimensional systems are the power-law type and the Berezinskii-Kosterlitz-Thouless (BKT) type. In fact, we explore the scaling as a function of $L/\xi$, where $\xi$ is the correlation length. We first consider $\xi$ as a power-law ansatz
\begin{eqnarray}
\xi_0 = \frac{1}{|\lambda - \lambda^*|^\nu}
\end{eqnarray}
with $\nu$ being the critical exponent and $\lambda^*$ is the disorder strength at the finite-size crossing point. For a finite-size system, the crossing point is a function of $L$; i.e. $\lambda^*(L)$. It has been observed numerically that the correlation length diverges at the phase transition \cite{Kjall_PRL:2014, Luitz_PRB:2015, Bertrand_PRB:2016, Luitz_PRB:2016}. However, recent phenomenological approaches with modified renormalization group schemes conjectured the BKT-type transition with correlation length \cite{Dumitrescu_PRB:2019, Goremykina_PRL:2019, Morningstar_PRB:2019, Suntajs_PRB:2020}
\begin{eqnarray}
\xi_{\rm BKT} = \exp\left[\frac{\nu}{\sqrt{|\lambda - \lambda^*|}}\right]
\end{eqnarray}
where $\nu$ is an unknown parameter. A comparison between these two types of correlation lengths has been numerically studied for uncorrelated disordered spin systems \cite{Suntajs_PRB:2020} by finding the best data collapse of average adjacent gap ratio and half-chain entanglement entropy. Here, we intend to analyze and compare the data collapse of $\langle r \rangle$ and $\mathcal{F}$ as a function of $L/\xi$. 

In general, the cost function for a set of $N_p$ data points corresponding to a generic observable $\mathcal{X}$ is given by
\begin{eqnarray}
\mathcal{C}_{\mathcal{X}} = \frac{\sum^{N_p -1}_{j = 1} |\mathcal{X}_{j+1} - \mathcal{X}_{j}|}{{\rm max}\{\mathcal{X}_j\} - {\rm min}\{\mathcal{X}_j\}} - 1.
\label{eq.costfun}
\end{eqnarray}
Here, $\mathcal{X}$ corresponds to either of the scaled fidelity susceptibility ($\mathcal{F}$) or $\langle r \rangle$. The standard approach is first to sort the $N_p$ values of $\mathcal{X}_j$ in nondecreasing order of $\Theta = {\rm sgn}[\lambda -\lambda^*]L/\xi$. In our calculation, we have considered $N_p \gtrsim 240$ for even numbers of lattice sizes $L= 12, 14, 16$ and $18$. In the event of a perfect data collapse, $\mathcal{C}_{\mathcal{X}} = 0$. However, for a reasonably large number of data sets as considered in this study, $\mathcal{C}_{\mathcal{X}}$ is always found to be positive. The idea is to find a best fit for which $\mathcal{C}_{\mathcal{X}}$ is minimum. In the next subsection, we extensively analyze the $\mathcal{C}_{\mathcal{X}}$ value by examining different ansatz for the finite-size crossing point and the correlation lengths.  

\subsection{Finite-size scaling}\label{sec.5}

We now aim to analyze the results of fidelity susceptibility $\mathcal{F}$ in terms of cost function minimization and compare them with the results of mean adjacent gap ratio $\langle r\rangle$ around the crossing point. At first, we consider two simplest functional forms of the disorder strength at the crossing point, one with a constant value $\lambda^* = \lambda_0$ and the other with a linear drift with system size $\lambda^* = \lambda_0 + \lambda_1L$. Here $\lambda_0$ and $\lambda_1$ are unknown parameters to minimize $\mathcal{C}_{\mathcal{X}}$.  

\begin{figure}[h]
\includegraphics[width=8.5cm,height=2.5cm]{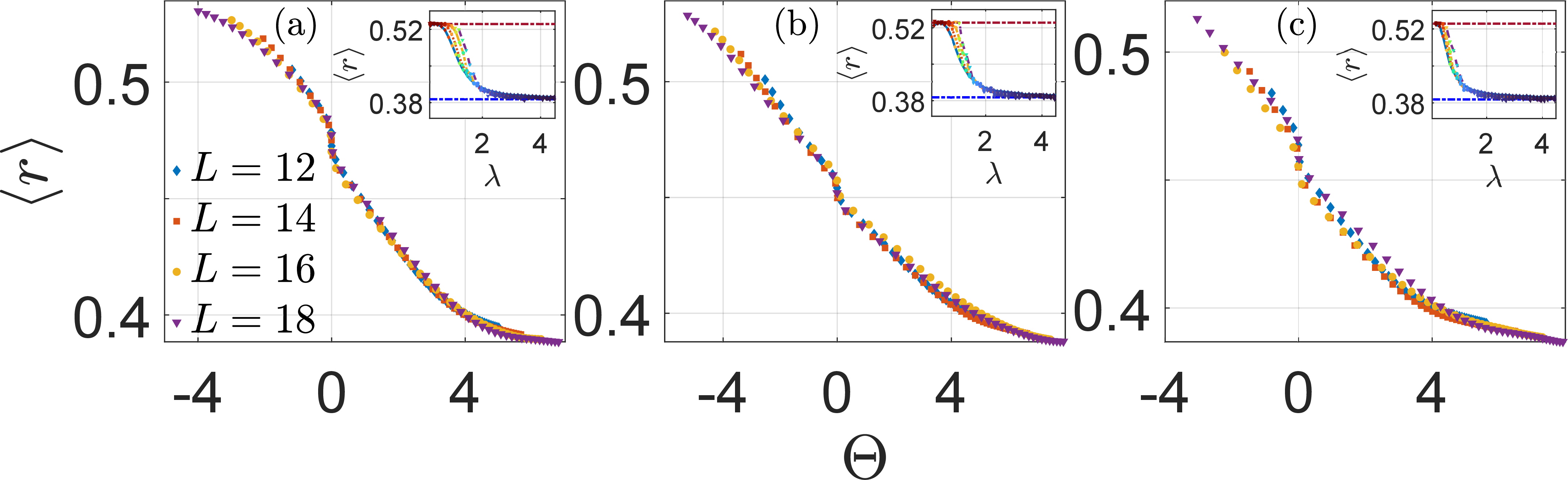}
\caption{Mean level spacing ratio $\langle r \rangle$ for different systems sizes $L$, calculated for $\alpha = 0$ (a), $0.3$ (b) and $0.7$ (c). Insets show $\langle r \rangle$ as a function of $\lambda$. In the main panels, we plot $\langle r \rangle$ as a function of the scaling parameter $\Theta= {\rm sgn}[\lambda -\lambda^*]L/\xi$, with $\xi = \xi_{\rm BKT}$ being a BKT correlation length, assuming the crossing point ansatz $\lambda^* = \lambda_0 + \lambda_1L$. The optimal parameters $\nu, \lambda_0$ and $\lambda_1$ in $\xi_{\rm BKT}$ are obtained by minimizing the cost function $\mathcal{C}_r$. The number of data points included in the minimization
procedure is $N_p = 264$.}
\label{fig 7}
\end{figure}
\begin{table}[h]
\begin{tabular}{c c c c c c} 
 \hline
  \textbf{$\alpha $}\hspace{0.15in} &  & $\lambda^* = \lambda_0$ \hspace{0.08in} & $\lambda_0 + \lambda_1 L$ \hspace{0.08in} &  $\lambda_0+\frac{\lambda_1}{L}$ \hspace{0.08in} & $\lambda_0+\frac{\lambda_1}{\textrm{ln}(L)}$ \\ [0.5ex] 
 \hline\hline
 \multirow{2}{*}{\textbf{$0.0$}}& $\mathcal{C}_r(\xi_0)$ & 2.9724 & 1.8717 & 2.8301 & 2.8781 \\[0.5ex] 
  & $\mathcal{C}_r(\xi_{\rm BKT})$ & 3.5283 & {\bf0.6994} & 1.9051 & 1.5855 \\ 
 \cline{1-6}
  \multirow{2}{*}{\textbf{$ 0.3$}}& $\mathcal{C}_r(\xi_0)$ & 2.2736 & 2.0104 & 2.2482 & 2.2586 \\ [0.5ex]
  & $\mathcal{C}_r(\xi_{\rm BKT})$ & 2.6288 & {\bf1.1197} & 1.2760 & 1.3502 \\
 \cline{1-6}
   \multirow{2}{*}{\textbf{$0.7$}}& $\mathcal{C}_r(\xi_0)$ & 3.7290 & 2.7762 & 3.4348 & 3.5133 \\ [0.5ex]
  & $\mathcal{C}_r(\xi_{\rm BKT})$ & 3.1716 & {\bf1.7609} & 2.4402 & 2.3607 \\  
 \cline{1-6}
\end{tabular}
\caption{Cost function $\mathcal{C}_r$ for the average adjacent gap ratio with $\alpha = 0, 0.3$ and $0.7$ using correlation length $\xi_0$ and $\xi_{\rm BKT}$. The columns denote different functional forms of $\lambda^*$ discussed in the text. The minimum values of $\mathcal{C}_r$, for different $\alpha$ values, are highlighted in bold font.}
\label{Table1}
\end{table}

In Fig. \ref{fig 7}, we plot the average adjacent gap ratio (between $\lambda_{\rm min} = 0.5$ and $\lambda_{\rm max} = 3.0$) as a function of $\Theta = {\rm sgn}[\lambda-\lambda^*]L/\xi$ with $\xi=\xi_{\rm BKT}$ and a linear drift of  the finite-size crossing point with system size $\lambda^* = \lambda_0 + \lambda_1L$ considering $L= 12, 14, 16$ and $18$. The insets represent the variation of $\langle r \rangle$ with respect to $\lambda$. To calculate $\langle r \rangle$ with $L = 12, 14$ and $16$, we adopt the energy window approach after an exact diagonalization as mentioned in Sec. \ref{sec.NuAp}. However, for $L = 18$, we use the Chebyshev filtering diagonalization to calculate $\langle r \rangle$ for 20 realizations. For different values of $\alpha = 0$, $0.3$ and $0.7$, we observe the signature of scaling collapse for the BKT-type correlation ansatz. At $\Theta = 0$, the collapse shows a discontinuous jump as a consequence of the singular behavior of $\xi_{\rm BKT}$ at the finite-size transition point $\lambda^*$. This kind of signature of data collapse of $\langle r\rangle$ has been previously observed for interacting disordered \cite{Khemani_PRX:2017, Dumitrescu_PRL:2017, Suntajs_PRB:2020} as well as quasi-periodic \cite{Aramthottil_PRB:2021} spin models. Similar to the spin models, we observe that the data collapse is more robust with functional dependence of $\lambda^*$ on $L$ as compared to a constant value $\lambda^* = \lambda_0$. In Table \ref{Table1}, we illustrate the results of data collapse $\mathcal{C}_r$ for power-law ($\xi_0$) and BKT-type ($\xi_{\rm BKT}$) correlation ansatz for different values of $\alpha$ (for details see Appendix \ref{Appendix-A}). 

\subsection{Outline}

We note that for all the cases the BKT-type correlation gives a better fit with a linear dependence of $\lambda^*$ on system size as observed in Ref. \cite{Aramthottil_PRB:2021}. As the value of $\alpha$ increases, the critical value $\lambda^*$ decreases for all the system sizes, which is evident from the phase diagram in Fig. \ref{fig 2}. Consistent with the earlier observation with quasi-periodic potential \cite{Khemani_PRL:2017}, the critical exponent for adjacent level spacing $\nu_{r}\sim0.5$ strongly violates the Harris-Luck criteria \cite{Luck_EPL:1993}. We further observe that the critical value $\lambda^*$ has a slope as high as $0.06$ for $\alpha = 0$ and $\sim 0.04$ for $\alpha = 0.3$ and $0.7$. Such system-size dependence naturally raises questions about the behavior of the critical value in the thermodynamic limit. This has been argued in Ref.\cite{Suntajs_PRB:2020}. To quantify the critical behavior of a disordered spin system in the thermodynamic limit, a generic functional form of the crossing point has been considered. Although this reduces the cost function to some extent, a significant system-size dependence for small values of $L$ still persists.

\begin{figure}[t]
\includegraphics[width=8.5cm,height=4.2cm]{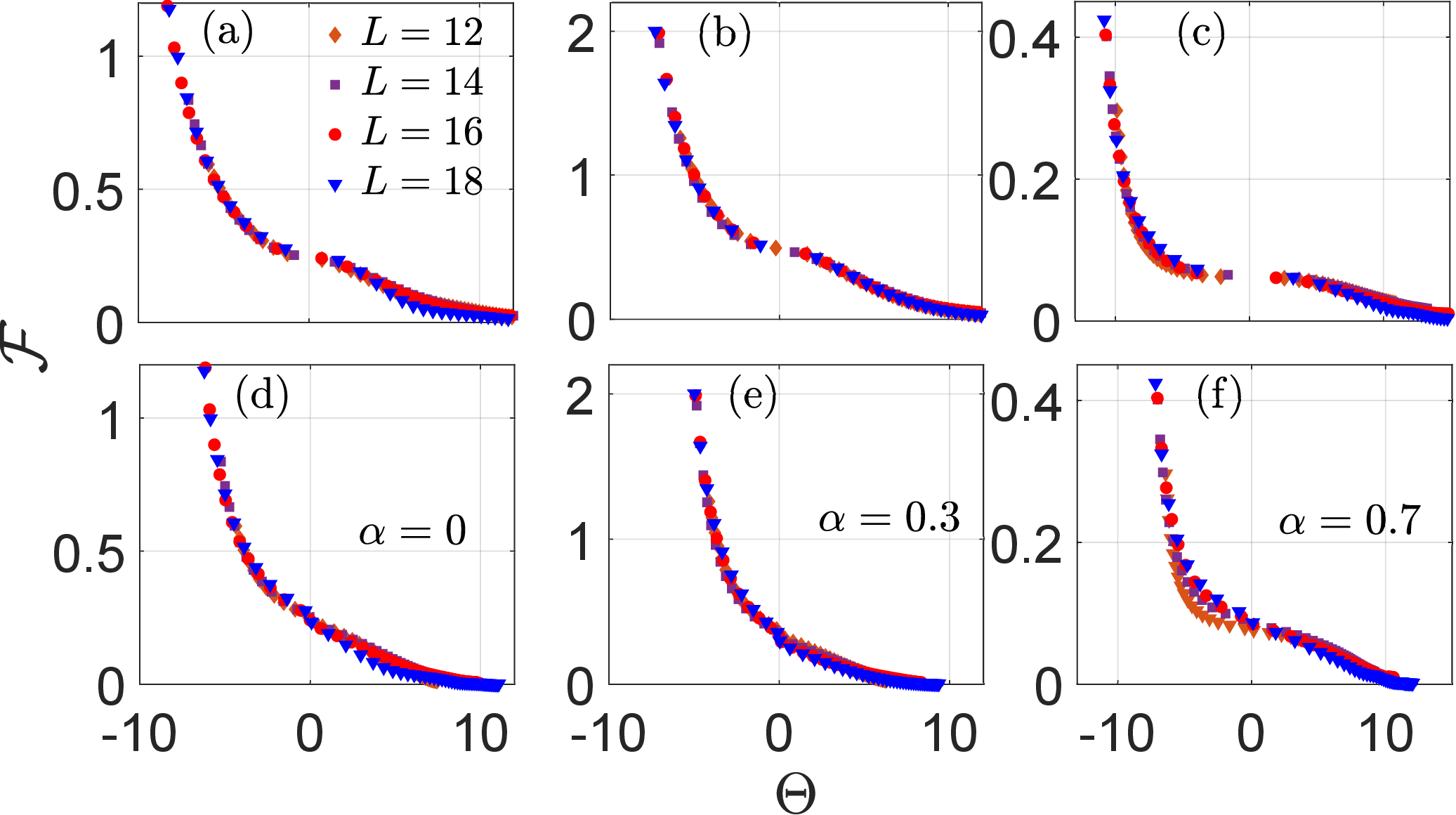}
\caption{Scaled fidelity susceptibility for extended operator and different system sizes are calculated for $\alpha = 0, 0.3$ and $0.7$, and plotted as a function of $\Theta= {\rm sgn}[\lambda -\lambda^*]L/\xi$, with $\xi = \xi_0$ (upper panel) and $\xi = \xi_{\rm BKT}$ (lower panel), assuming the crossing point ansatz $\lambda^* = \lambda_0 + \lambda_1L$. The parameters $\nu, \lambda_0$ and $\lambda_1$ in $\xi_{\rm BKT}$ are obtained by minimizing $\mathcal{C}_\mathcal{F}$. The dotted vertical line represents the critical point $\lambda=\lambda^*$ at which $\mathcal{F}$ becomes discontinuous. The number of data points included in the minimization procedure is $N_p = 240$.}
\label{fig 8}
\end{figure}
\begin{table}[h]
\begin{tabular}{c c c c c c} 
 \hline
  \textbf{$\alpha $}\hspace{0.15in} &  & $\lambda^* = \lambda_0$ \hspace{0.08in} & $\lambda_0 + \lambda_1 L$ \hspace{0.08in} &  $\lambda_0+\frac{\lambda_1}{L}$ \hspace{0.08in} & $\lambda_0+\frac{\lambda_1}{\textrm{ln}(L)}$ \\ [0.5ex] 
 \hline\hline
 \multirow{2}{*}{\textbf{$0.0$}}& $\mathcal{C}_\mathcal{F}(\xi_0)$ & 0.6501 & {\bf 0.5970} & 0.6340 & 0.6348 \\[0.5ex] 
  & $\mathcal{C}_\mathcal{F}(\xi_{\rm BKT})$ & 0.8629 & 0.7694 & 0.8584 & 0.8540 \\ 
 \cline{1-6}
  \multirow{2}{*}{\textbf{$ 0.3$}}& $\mathcal{C}_\mathcal{F}(\xi_0)$ & 0.2549 & {\bf 0.2199} & 0.2401 & 0.2313 \\ [0.5ex]
  & $\mathcal{C}_\mathcal{F}(\xi_{\rm BKT})$ & 0.4115 & 0.3729 & 0.3889 & 0.4061 \\
 \cline{1-6}
   \multirow{2}{*}{\textbf{$0.7$}}& $\mathcal{C}_\mathcal{F}(\xi_0)$ & 1.8233 & {\bf 0.7263} & 1.4552 & 1.1018 \\ [0.5ex]
  & $\mathcal{C}_\mathcal{F}(\xi_{\rm BKT})$ & 1.5370 & 1.0088 & 1.3807 & 1.2401 \\  
 \cline{1-6}
\end{tabular}
\caption{$\mathcal{C}_\mathcal{F}$ for the scaled fidelity susceptibility for extended operator with $\alpha = 0, 0.3$ and $0.7$ using correlation length $\xi_0$ and $\xi_{\rm BKT}$. The optimal values are shown in bold font.}
\label{Table2}
\end{table}

We now move to discuss the finite-size scaling behavior of the scaled fidelity susceptibility. In Fig. \ref{fig 8}, we display the cost-optimised fidelity susceptibility for the extensive operator with different (even) system sizes $L = 12, 14, 16$ and $18$ for $\alpha = 0, 0.3$ and $0.7$ near the crossing point. In Figs. \ref{fig 8} (a, b, c), we have shown the power law ansatz of the correlation length $\xi = \xi_0$ and in Figs. \ref{fig 8} (d, e, f) the correlation length is BKT-type; i.e., $\xi = \xi_{\rm BKT}$. Here, we consider only even numbers of lattice sizes, as the corresponding cost function is minimum. In Appendix \ref{Appendix-B}, we show the optimal cost function considering both even and odd numbers of lattice sizes for the local perturbation.  In Table \ref{Table2}, we illustrate $\mathcal{C}_{\mathcal F}$ quantitatively for different values of $\alpha$ with the system-size dependence as realized for the $\mathcal{C}_r$ calculation. The cost-minimized $\mathcal{F}$ as a function of $\Theta$ with $\xi_{\rm BKT}$ for different types of functional scaling is presented in Appendix \ref{Appendix-C}. For this particular case we consider $N_p = 240$ for all $\alpha$ values. However, the regions of $\lambda$ through which $\mathcal{C}_{\mathcal F}$ is minimized for different $\alpha$ values are considered to be different (see Appendix \ref{Appendix-A}). Similar to the $\langle r \rangle$ calculation, we use the energy window approach for $L \leq 16$. We calculated $\mathcal F$ only near the crossing region using the 1500 interior eigenstates obtained from the Chebyshev filtering. We find that, unlike the earlier case where $\mathcal{C}_r$ is optimum for BKT-type correlation ($\xi = \xi_{\rm BKT}$), $\mathcal{C}_{\mathcal F}$ is minimum for a power-law correlation ($\xi = \xi_0$) with a critical value having linear system-size dependence. This contrasting scaling behavior is significant. This is generally interpreted by identifying the distinct nature of the two different identifiers. The adjacent gap ratio is sensitive to global spectral structure, while the fidelity susceptibility is associated with the local integrals of motion. Near the MBL transition, the level statistics may be strongly influenced by rare thermal inclusions and thermal avalanches, both of which have been predicted to demonstrate BKT-like scaling \cite{Goremykina_PRL:2019, Dumitrescu_PRB:2019, Suntajs_PRB:2020}. Conversely, the fidelity susceptibility is sensitive to eigenstate structure in addition to eigenvalue statistics and therefore couples to local critical fluctuations. In the thermodynamic limit, both local and global properties should have similar behavior. The discrepancy between the two different scaling correlations is an indication of a finite-size effect that leads to a crossover rather than a distinct transition.
\begin{figure}[h]
\includegraphics[width=8.5cm,height=4.2cm]{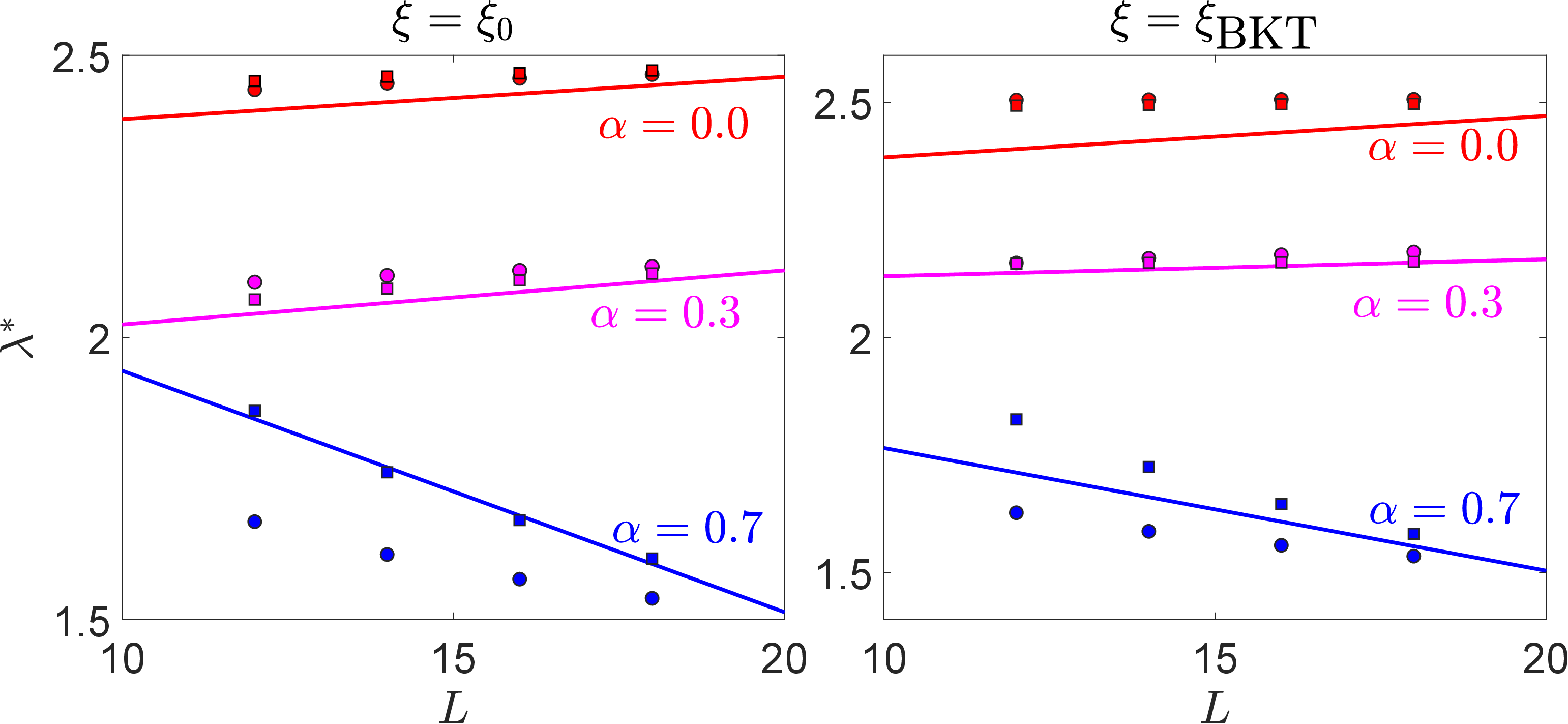}
\caption{Critical point $\lambda^*$ as a function of system size $L$, for the power-law correlation $\xi =\xi_0$ (left) and BKT-type correlation $\xi = \xi_\textrm{BKT}$ (right) for different values of $\alpha$. $\lambda^*$ are obtained from the best data collapse of  $\mathcal{C_{\mathcal{F}}}$. The straight lines represent the linear dependence of the critical value $\lambda^* = \lambda_0 + \lambda_1L$. The square and circular markers are different nonlinear ansatz $\lambda^* = \lambda_0 + \lambda_1/\textrm{ln}(L)$ and $\lambda^* = \lambda_0 + \lambda_1/L$, respectively.}
\label{fig 9}
\end{figure}

In Fig. \ref{fig 9}, we display the system-size dependence on the critical value $\lambda^*$ for different functional forms as discussed above. For both power-law and BKT-type correlation, we obtained the critical exponent $\nu_{\mathcal F} \sim 0.5$ for $\alpha =0$ and $0.3$, and $\nu_{\mathcal F }\sim 0.4$ for $\alpha =0.7$, similar to $\nu_r$. Although the scaling exponent exhibits a robust behavior across the finite system sizes considered, these results alone are insufficient to conclusively determine the asymptotic universality class in the thermodynamic limit.
It is observed that for $\alpha = 0$ and $0.3$, $\lambda^*$ has a slow positive drift with system size for both power-law and BKT-type ansatz. However, for $\alpha = 0.7$ we observe a comparatively rapid negative drifting of $\lambda^*$ with system size. This unusual behavior in the large $\alpha$ limit may be interpreted by identifying the fact that with increasing $\alpha$, the ergodic region shrinks and for $\alpha = 0.7$, the effect of the mobility edge may cause instability in the ergodic-to-localized transition. Such behavior of $\lambda^*$ is mainly attributed to the mixing of more localized single-particle orbitals with increasing $\alpha$ in the presence of interaction \cite{Li_PRL:2015}. The anomalous negative drifting of $\lambda^*$ with $L$ and a lower scaling exponent $\nu_\mathcal{F} \sim 0.4$ are manifestations of this many-body mobility edge effect at higher values of $\alpha$. This may be overcome by considering a few larger system sizes ($L > 18$) and a larger number of realizations for $L \geq 18$. Furthermore, it is crucial to note that while the slope of linear drift with system size $L\leq 18$ for $\langle r \rangle$ is $\lambda_1 \sim 0.06$ for $\alpha = 0$ and $\sim 0.04$ for $\alpha = 0.3$, for $\mathcal{F}$, the coefficients of linear drifting are $\lambda_1 = 0.0075$ and $0.0036$ for $\alpha =0$ and $0.3$, respectively. Albeit small, the linear dependence of $\lambda^*$ on $L$ may not be accurately extrapolated to determine the critical disorder in the thermodynamic limit. The possible ways to reduce these uncertainties are: (i) by considering a multi-parameter fitting ansatz for $\lambda^*(L)$; (ii) a few more larger system-size considerations; and (iii) by enhancing the number of eigenstates inside a particular energy window used to calculate fidelity susceptibility with a Chebyshev filtering process for larger system sizes ($L>16$). Nevertheless, this particular signature may be regarded as the small-size behavior of critical crossing. A suppressed linear drift affirms the sensitivity of the fidelity susceptibility even in small-size systems, essentially paving the way to estimate the critical value in the thermodynamic limit.

\section{Conclusion}\label{sec.6}

We investigated the thermalization properties of a quasi-periodic interacting generalized Aubry-Andr\'e model. By analyzing adjacent gap ratio statistics, we identified the ergodic-to-localized transition region for finite-sized systems. We further examined the spectral form factor to quantitatively characterize the Thouless time, which governs the onset of thermalization. Our results reveal that the thermalization behavior is influenced by both the generalized parameter $\alpha$ and the quasi-periodic potential strength $\lambda$. To deepen our understanding of the ergodic-localized phase transition and determine the critical parameters, we analyzed the fidelity susceptibility across different system sizes. The susceptibility exhibits a peak near the finite-size transition point, with the peak value drifting as the system size increases, displaying behavior similar to that observed in disordered spin systems \cite{Sels_PRE:2021}. Since the intermediate regimes in the fidelity susceptibility phase diagrams show an enhanced sensitive region to both local and global perturbation with a maximum chaotic point, the many-body eigenstates will be extended. However, accessing the multifractal behavior in this regime requires rigorous analysis of the fractal dimension associated with the many-body eigenstates in the presence of interaction compared to the non-interacting case. Nevertheless, due to the presence of a mobility edge in the considered model, the eigenstates in this regime are expected to show some multifractal characters.

Furthermore, we optimized the cost function to observe the finite-size scaling behavior. We noticed that while the optimization of the adjacent gap ratio shows a better fit for the BKT-type correlation length with a linear drift of finite-size crossing point, the cost function minimization for the fidelity susceptibility gives a better data collapse for the power-law correlation length with linear drift. The contrasting scaling behaviors are attributed to finite-size effects rather than regarded as universal features, suggesting a crossover rather than a sharp transition. The AGP prescription resembles a remarkable reduction in the system-size dependence on the critical value, at least for the lower values of $\alpha$, as compared to other quantifiers. With the availability of limited computational resources, we have gone up to $L=18$ system sizes to capture the finite-size characteristics of the considered model. As an outlook, one may consider larger system sizes and make a rigorous analysis of AGP to scrutinize the stability of the critical disorder strength in the thermodynamic limit, particularly for the larger value of the GAA parameter.  

\section*{Acknowledgment}

Calculations reported in this work were performed on the ParamVikram-1000 HPC cluster at the Physical Research Laboratory (PRL), Ahmedabad, Gujarat, India. SM and BKS would like to thank the Department of Space, Government of India, for the support to carry out this work.

\appendix
\renewcommand{\thesection}{\Alph{section}} 

\section{Cost function minimization protocol}\label{Appendix-A}
Here, we describe the minimization protocol of the cost function of Eq. (\ref{eq.costfun}) in detail. We first calculate the eigenstates of the GAA Hamiltonian by exact diagonalisation and calculate the observables such as the adjacent gap ratio $\langle r\rangle$ and fidelity susceptibility $\mathcal{F}$. We then take the average over the number of eigenstates and different realizations, considering random offset values of $\phi$ for each realization. After identifying the crossing region, we consider a significant amount of data in that region. Then, the data is arranged in a non-descending order of the scaling parameter $\Theta = {\rm sgn}[\lambda - \lambda^*]L/\xi$. For different ansatz of correlation lengths, we minimize the cost function by optimizing the set of parameters. We use the differential evolution method obtained from the Metaheuristics library in the Julia programming package for the optimization. For our study we have used two types of correlation length ansatz $\xi_0$ and $\xi_{\rm BKT}$, and four different functional forms of critical disordered strength $\lambda^*$ with respect to system size. For Fig. \ref{fig 7} and Table \ref{Table1}, we have used $N_p = 264$ data points in the region $\lambda\geq 1$. The optimized values of $\mathcal{C}_r$ for different values of $\alpha$ depend on the number of data points we consider as well as the region of $\lambda$ around the crossing point. 

In the case of cost function optimization of the fidelity susceptibility, as the crossing point shifts with different values of $\alpha$, we choose three different ranges of $\lambda$ for $\alpha = 0, 0.3$ and $0.7$ to optimize the $\mathcal C_{\mathcal F}$. For $\alpha = 0, 0.3$ and $0.7$, the ranges of $\lambda$ are considered to be $[2.2, 3.6]$, $[1.75, 3.15]$ and $[1.3, 2.7]$, respectively.  

\section{Finite-size behavior of fidelity susceptibility for local operator}\label{Appendix-B}

In the main text, we have only discussed the scaling behavior of the fidelity susceptibility for an extensive operator with an even number of lattice sites. Here, we discuss the finite-size scaling of fidelity susceptibility for the local operator $H_{1} = n_{L/2}$ with $L$ even and for an odd number of lattice sites, we consider $H_{1} = n_{(L+1)/2}$. We calculate the optimal $\mathcal{C}_{\mathcal F}$ considering $L= 13, 14, 15$ and $16$ as can be found in Table \ref{Table3}. The calculated values of $\mathcal{C}_{\mathcal F}$ for both power-law and BKT-type correlation are significantly larger when incorporating both odd and even lattice sizes as compared to the case with an even number of system sizes as given in Table \ref{Table2}.

\begin{table}[h]
\begin{tabular}{c c c c c c} 
 \hline
  \textbf{$\alpha $}\hspace{0.15in} &  & $\lambda^* = \lambda_0$ \hspace{0.08in} & $\lambda_0 + \lambda_1 L$ \hspace{0.08in} &  $\lambda_0+\frac{\lambda_1}{L}$ \hspace{0.08in} & $\lambda_0+\frac{\lambda_1}{\textrm{ln}(L)}$ \\ [0.5ex] 
 \hline\hline
 \multirow{2}{*}{\textbf{$0.0$}}& $\mathcal{C}_\mathcal{F}(\xi_0)$ & 1.7846 & 1.6900 & 1.7642 & 1.7706 \\[0.5ex] 
  & $\mathcal{C}_\mathcal{F}(\xi_{\rm BKT})$ & 1.9160 & 1.6957 & 1.9008 & 1.9102 \\ 
 \cline{1-6}
  \multirow{2}{*}{\textbf{$ 0.3$}}& $\mathcal{C}_\mathcal{F}(\xi_0)$ & 1.8753 & 1.2944 & 1.7285 & 1.7359 \\ [0.5ex]
  & $\mathcal{C}_\mathcal{F}(\xi_{\rm BKT})$ & 1.7520 & 1.5007 & 1.8717 & 1.8751 \\
 \cline{1-6}
   \multirow{2}{*}{\textbf{$0.7$}}& $\mathcal{C}_r(\xi_0)$ & 2.0831 & 2.0667 & 2.0766 & 2.0806 \\ [0.5ex]
  & $\mathcal{C}_\mathcal{F}(\xi_{\rm BKT})$ & 2.1944 & 2.1469 & 2.2615 & 2.2576 \\  
 \cline{1-6}
\end{tabular}
\caption{Optimal cost function of the scaled fidelity susceptibility for local operator with different values of $\alpha$ using correlation length $\xi_0$ and $\xi_{\rm BKT}$.}
\label{Table3}
\end{table}

\section{Various critical scaling}\label{Appendix-C}

\begin{figure}[h]
\includegraphics[width=8.5cm,height=4.5cm]{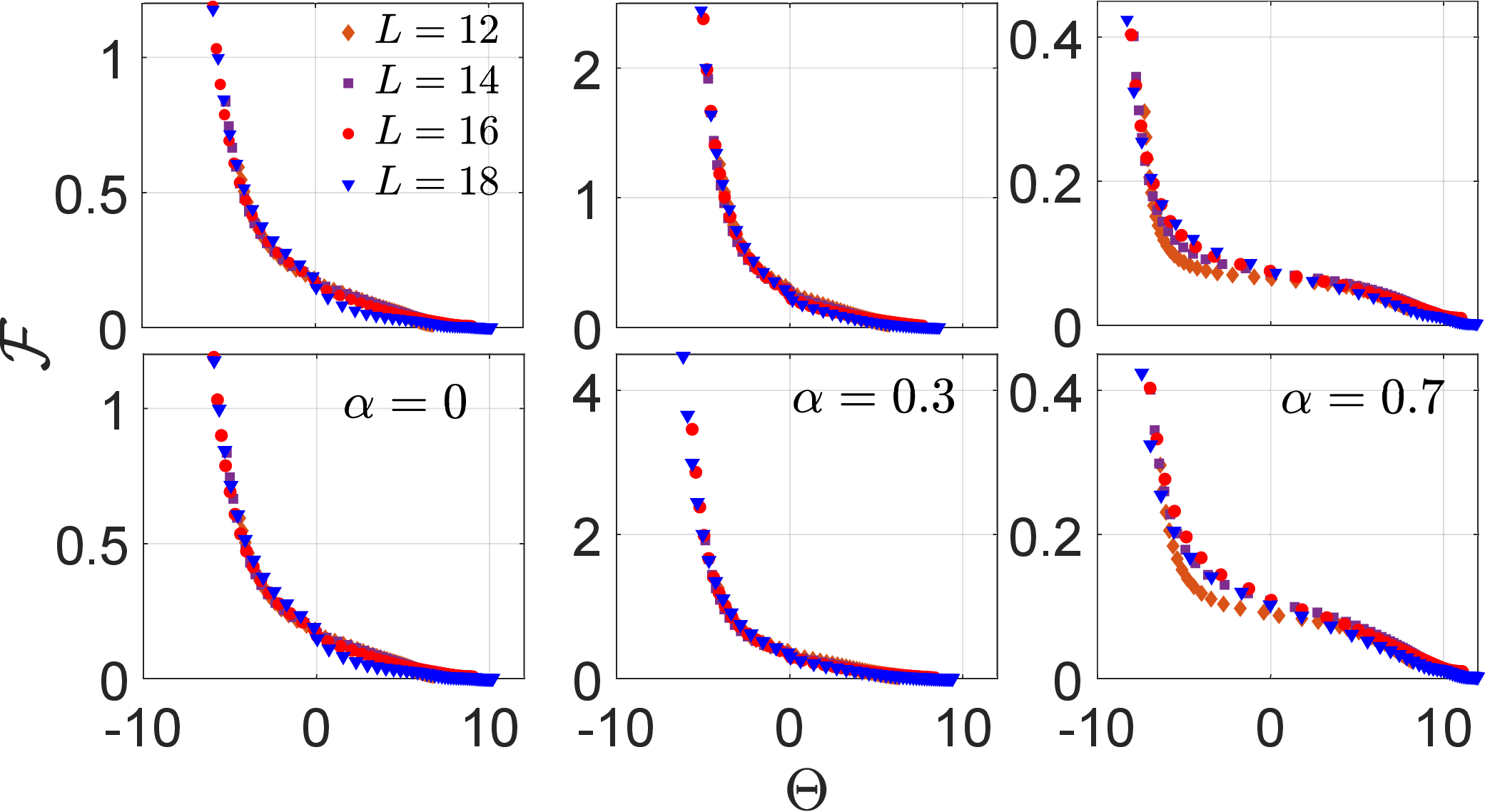}
\caption{$\mathcal{F}$ for the extended operator are plotted with $\xi = \xi_{\rm BKT}$, assuming the critical ansatz $\lambda^* = \lambda_0 + \lambda_1/{\rm ln}(L)$ (upper panel) and $\lambda^* = \lambda_0 + \lambda_1/L$ (lower panel).}
\label{fig 11}
\end{figure}

As discussed in Sec. \ref{sec.5}, we have considered four different types of scaling ansatz for the critical value $\lambda^*$. In Fig. \ref{fig 8} of the main text, we have shown the scaling collapse of fidelity susceptibility for the BKT-type ansatz with a critical value that has linear dependence on system size. Here, in Fig. \ref{fig 11}, we show the scaling collapse for BKT-type correlation with the two other functional forms of critical values having different nonlinear dependence on system size.

\bibliography{cite_MBL_GAAH}

\end{document}